%% file: main.tex
\documentclass[sigconf,nonacm]{acmart}
\usepackage{graphicx}
\usepackage{subcaption}
\usepackage{tabularx}
\usepackage{array}
\usepackage{booktabs}
\usepackage{rotating}
\usepackage{times}
\usepackage{url}
\usepackage{enumitem}
\usepackage{float}
\usepackage{placeins}
\usepackage{xspace}
\usepackage{hyperref}
\usepackage{afterpage}
\afterpage{\clearpage}
\usepackage{fancyhdr}

\newcommand{\systemone}{query scheduler\xspace}
\newcommand{\systemtwo}{query optimizer\xspace}

\begin{document}
\title{Multi-agent Databases via Independent Learning\\(Extended Abstract)}
\author{Chi Zhang}
\affiliation{%
  \institution{Brandeis University}
}
\email{chizhang@brandeis.edu}

\author{Olga Papaemmanouil}

\affiliation{%
  \institution{Brandeis University}
}
\email{opapaemm@brandeis.edu}

\author{Josiah P. Hanna}
\affiliation{%
  \institution{University of Wisconsin-Madison}
}
\email{jphanna@cs.wisc.edu}

\author{Aditya Akella }
\affiliation{%
  \institution{University of Texas, Austin}
}
\email{akella@cs.texas.edu}


\begin{abstract}
Machine learning is rapidly being used in database research to improve the effectiveness of numerous tasks including but not limited to query optimization, workload scheduling and physical design. Currently, the research focus has been on replacing a single database component responsible for one task with its learning-based counterpart. However, query performance is not simply determined by the performance of a single component but by the cooperation of multiple ones. As such, learning-based database components need to collaborate during both training and execution in order to develop policies that meet end performance goals. Thus, the paper attempts
to address the question  "\emph{Is it possible to design a database consisting of various learned components that cooperatively work to improve end-to-end query latency?}". 

To answer this question, we introduce MADB (Multi-Agent DB), a proof-of-concept system that incorporates a learned query scheduler and a learned query optimizer. MADB leverages a cooperative multi-agent reinforcement learning approach that allows the two components to exchange the context of their decisions with each other and collaboratively work towards reducing the query latency. Preliminary results demonstrate that MADB can outperform the non-cooperative integration of learned components.
\end{abstract}

\maketitle

\pagestyle{empty}
\input{intro}

\input{model}
\input{experiments}
\input{conclusion}

\bibliographystyle{abbrv}
\bibliography{sample}

\end{document}

%% file: intro.tex
\section{Introduction}
Database (DB) systems drive various user-facing applications and services and play a key role in user-perceived performance. Modern DBs have many components (optimizer, scheduler, physical designer, etc.), and machine learning is rapidly leveraged to replace numerous of these components by their learning-based counterparts~\cite{deep_card_est2,negi2020cost,yang2019deep,van2017automatic,marcusbenchmarking,sun2019end,yangneurocard,tzoumas2008reinforcement}. These ML-driven approaches have demonstrated great potential to improve query performance as well as automatically adapt to the dynamic environments modern data management applications often operate in. 

 Today, the bulk of work on learned databases focuses on replacing a \emph{single} component or task of the database with a learned-based solution, and these learned components are designed \emph{independently} of each other.  In reality, though, production database settings require careful optimizations across all these components as query performance is determined by the cooperation of these components. For example, the optimizer's data access decisions should be in coordination with the physical designer's re-distribution actions and the buffer pool manager's  caching decisions. However, with the replacement of these components with learned ones,  it is unclear if and how these per-component solutions can be integrated into the same database and how they can cooperatively yield end-to-end benefits. 
 
 In this paper, we argue that new database designs that optimize decisions across learned components (aka learning \emph{agents}) are fundamental to
good end-to-end query performance. We introduce, \emph{Multi-Agent DB} (MADB), to demonstrate the benefits of such across-component coordination. MADB includes two key reinforcement learning agents: a learned \systemone and a learned \systemtwo. Despite the fact that these two ML-driven systems operate in the same environment, they have different tasks and performance goals. Given a set of queries queued for execution, the \systemone agent strives to identify the query execution ordering that improves the data cache utilization. The objective of the \systemtwo agent is to identify the query execution plan that minimizes a given query's execution time. 

While both of these goals relate to query performance, without coordination, these components might miss opportunities for further performance improvements or might even make conflicting decisions. For instance, the scheduler learns to prioritize queries that reuse cached data by observing both the contents of the buffer pool and the expected execution plans of the queued queries. However,  query plan decisions (i.e., full scans vs. index scans, join orderings) are determined by the optimizer's model, which adjusts its policy based on the observed execution time of each plan. Execution times, though, depend on buffer pool hits which are determined by the data access methods used by the query plan. In a non-cooperative multi-agent setting, such interdependent decisions are not captured properly, which could lead to obsolete policies for all involved agents. 

To address this challenge, this paper attempts to address the question ``\emph{Is it possible to design a database consisting
of various learning components that cooperatively work to
improve end-to-end query performance?}``. We study this question by leveraging a cooperative multi-agent reinforcement learning (MARL) approach. MARL~\cite{tan1993multi,zhang2021multi} is a sub-field of reinforcement learning that studies how multiple reinforcement learning models (agents) interact in a common environment in order to accomplish a common goal. In particular, MADB relies on independent learning (IL)~\cite{de2020independent,lowe2017multi}, a MARL approach that distributes and decentralizes the construction of a system's agents (in our case, learned database components). Based on IL, the learned scheduler and learned optimizer of MADB are trained in the same environment, adjust their policies based on each other's observations, and improve end-to-end query performance, while maintaining their own performance objectives. 

We presented our system and formalized our learning task in Section~\ref{t:framework}. We present preliminary experimental results  from a proof-of-concept prototype implementation in Section~\ref{t:preliminary}, 
and in Section~\ref{sec:conclusion} we highlight directions for future work.

%% file: model.tex
\section{MADB Overview}\label{t:framework} 
This section describes MADB, a proof-of-concept multi-agent database that combines two learned components: a scheduler and an optimizer. We present these two models as non-cooperating agents and describe how they collaborate to improve query latency.
\subsection{Learned Query Scheduler} \label{qs}
MADB incorporates the SmartQueue learned scheduler ~\cite{zhang}, is a deep reinforcement learning scheduler that aims to improve a weighted average between short-term buffer hits and the long-term impact of query scheduling choices. Next, we briefly describe the scheduler and its inputs assuming no cooperation with the optimizer. More details can be found~\cite{zhang}.   

Our scheduler uses a  Q-learning approach \cite{deep_rl} to decide which query to execute next. It operates over a set of environmental states $S^{sche}$ (the contents in the buffer pool, query access patterns) and a set of actions $A^{sche}$ (candidate queries to execute next). The query scheduler models the scheduling problem as a Markov Decision Process (MDP)~\cite{rl_book}: by selecting one query from the queue to execute, the agent transitions from its current state to a new one. By executing a new query on the current state, the agent receives a reward calculated as the buffer hit ratio of the executed query. SmartQueue relies on a continuous learning process: as more queries arrive and are scheduled for execution by the model, the agent collects more information on the context of these decisions and their impact on its reward/ performance goal and adapts its policy accordingly. This allows SmartQueue to automatically adapt to dynamic workloads.


{\bf Model Input} One first input to the scheduler's model is the state of the buffer pool, namely which blocks are currently cached in
memory. This information is captured in a 2-dimensional bitmap $B$ where rows represent base relations and columns represent data blocks. The $(i,j)$ entry is 1 if the $j$-th block of relation $i$ is cached and 0 otherwise. Each row is downsized by computing a simple moving average over the number of its block elements. The second input feature is the expected data block requests of each queued query.  The scheduler collects the query's  plan and estimates the probability of each input table's data block being accessed. We also include index data access blocks when used by the query plan. The query vector is downsized in the same as the buffer bitmap.

\subsection{Learned Query Optimizer} \label{t:qo}

MADB's second learned component is a query optimizer. While a number of learning-based query optimization approaches have been proposed~\cite{rejoin, neo,wang2016database,krishnan2018learning,ortiz2018learning,trummer2018skinnerdb,marcus2020bao,negi2021steering}, our current learned optimizer is based on a variation of ReJoin~\cite{rejoin}, a deep reinforcement learning optimizer that decides the best join orderings of a query plan. Join ordering can have a significant impact on query execution, and using deep reinforcement learning  has already been proven effective~\cite{rejoin}.

In contrast to ReJoin, our optimizer is trained by learning from demonstrations~\cite{schaarschmidt2018lift,marcus2018towards,neo}, which builds a model in two steps. The first phase allows the learned agent to understand how a traditional query optimizer optimizes a query (i.e., picks the join orderings in our case). This method begins by training an agent to mimic the behavior of a traditional optimizer (aka expert) and alleviate the cold start problem of machine learning algorithms. 

The model uses as a Markov Decision Process (MDP)~\cite{rl_book} to decide the orders input relations will be joined.  When a query is sent by the scheduler, the optimizer takes an action: it chooses which base relation to join next and moves from the current state to a new  one that includes a new join. Given this join ordering, the optimizer adds to it information about data access operators (e.g., index scans), and the plan is executed. The execution time of each join is recorded and its latency is used as the reward for the action that added that join. The process continues until all relations to be joined.

This agent is first trained on a representative query set and learns to mimic the traditional optimizer's join ordering decisions. The final model is then used used during runtime to decide the ordering of queries. This runtime reinforcement learning model also adapts over time: as more queries are executed, the Q-learning policy of the optimizer is continuously fine-tuned and customized automatically to dynamic workloads. More details can be found~\cite{neo, marcus2018towards}.

{\bf Model Input} The first input for the \systemtwo is the join ordering of the \emph{latest} executed query. This state is represented as a matrix with rows for base relations and columns for joins.
 If base relation $A$ is to be joined with base relation $B$ and that join is in the $k$-level of the join tree, the matrix values of both of these relations are set to k. Low-level base relations are joined first, then high-level base relations. The second input of the optimizer is a vector indicating the base relations to be joined in the current query. 
 
\begin{figure}
  \centering
  \includegraphics[width=0.5\textwidth]{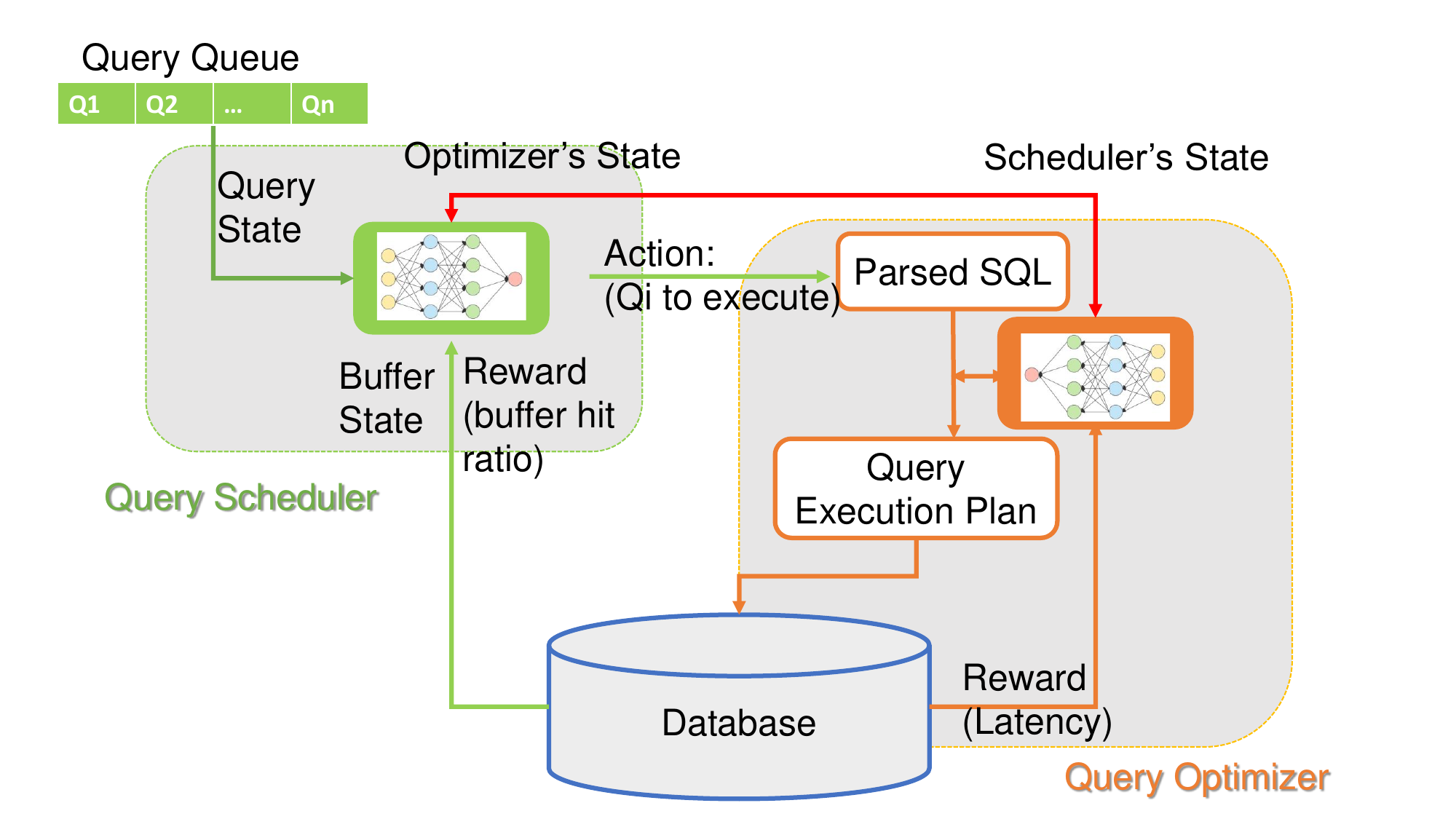}
  \caption{MADB System Model}
  \label{fig:system_model}
  \vspace{-4mm}
\end{figure}

\subsection{Cooperative learning in MADB}\label{s:coop}
In this section, we describe how the scheduler and the optimizer agents can effectively cooperate in the same environment.   
We use a 
multi-agent reinforcement learning (MARL) approach that requires that all agents in the environment coordinate and collaborate in order to accomplish a common goal~\cite{tan1993multi,zhang2021multi,de2020independent}. 

Different types of coordination and collaboration approaches have been proposed for  MARL techniques~\cite{zhang2021multi}. In this work, we chose to capture the interaction between the scheduler and the optimizer using an independent learning (IL) approach~\cite{de2020independent}. In IL, the MARL problem is decomposed into a problem of multiple independent agents, in which all other agents are considered to be part of the environment, and an agent's learning policy is determined solely by the agent's local observation history. More specifically, we use the popular paradigm of centralized-training-decentralized-execution~\cite{de2020independent,lowe2017multi} in which each agent learns a policy based on information they observe at execution time but uses shared information during training.

In our setting, the \systemone and the \systemtwo are trained concurrently with the goal of lowering the overall execution time of query workload. However, they achieve this goal by decomposing this central goal into two decentralized problems, with the scheduler focusing on improving the average buffer hit ratio and the optimizer focusing on reducing the latency of a single query. Figure \ref{fig:system_model} depicts how these agents observe the environment's state, collaborate, and share information to ultimately accomplish the common goal of reducing the total query latency of the query workload.

Let us assume there are N queries in the scheduler's queue ($Q_1$,...$Q_n$.). Once the scheduler decides a query $Q_i$ to execute, it sends it to the optimizer to generate its execution plan. Each agent will make its decisions based on its  past observations. In our cooperative setting, these observations include information from the other agent. That way, each agent is seen as part of the other agent's environment and their  respective states and actions will influence the formulation of the other's policies. 


Specifically, at each step $t$ the \systemone observes (a) the status of the buffer pool, {$BP_t^{sche}$}, i.e.,  which data blocks are cached, and (b) the expected data block requests, {$BR_t^{sche}$} of queries in the queue and (c) the optimizer's state. This state includes an encoding of the last executed query plan, {$QEP_t^{opt}$}, and an encoding of the join orderings, {$JD_t^{opt}$}, sent by the optimizer for each query $Q_i$ it receives for execution.  The encoding of the execution plan is similar to the one used by ReJOIN~\cite{rejoin}. Join ordering is encoded by assigning a value $k$ to all joined base relations, where $k$ denotes the level that the base relation appears in the join query plan.

The \systemone's state is also shared with the optimizer. This state includes the encoding of the buffer contents, {$BP_t^{sche}$}, as well as the encoding of the expected data blocks to be accessed by the query $Q_i$, {$BR_t^{sche}$}. This state is added to the optimizer's state (the encoded join ordering $JD_t^{opt}$ and the last executed query plan, {$QEP_t^{opt}$}, and the optimizer decides the best join ordering for $Q_i$. The database  executes the final query plan and its latency {$R_t^{LAT}$} and average buffer hit ratio {$R_t^{BHR}$} are collected as the reward of the optimizer and scheduler respectively. This process of training is continuous: as queries enter the system, both agents adjust their observations as well as their actions and policies.

The scheduling policy of the \systemone is expressed as a Q-learning reward function {$Q_(S_t^{sche},A_t^{sche})$}, that outputs a Q-value for taking an action {$A_t^{sche}$} (i.e., a query to execute next) on a current state {$S_t^{sche}$}. The Q-value {$Q(S_t^{sche}, A_t^{sche})$} is calculated by adding the maximum reward from future states to the reward for obtaining the present state, effectively influencing the current scheduling decision. This potential reward is a weighted average of all future buffer hit ratios starting from the current state. The agent learns a new policy {$Q^{new}(S_t^{sche}, A_t^{sche})$} after each action {$A_t^{sche}$} on a state {$S_t^{sche}$}:
\begin{equation*}
Q\left(S_t^{sche}, A_t^{sche}\right)+\alpha\left[R_t^{BHR}+\gamma \max _{\alpha}\left(Q\left(S^{sche}_{t+1}, \alpha\right)-Q\left(S_t^{sche}, A_t^{sche}\right)\right)\right]
\end{equation*}
\noindent where $S_t^{sche}$ is the encoded scheduler's state. 


The learning policy of the \systemtwo is to decide which base relation to join next. After joining all relevant base relations, a final query execution plan is produced. When a query is finished, the agent takes an action $A_t^{opt}$ and shows which base relation to be joined next based on the current state $S_t^{opt}$. The reward $R_t^{LAT}$ for the action $A_t^{opt}$ is calculated based on the query latency $LAT$. Thus, we will train and learn the policy that maximizes its total reward $Q_t^{opt}$. At each step t, the agent updates $Q^{new}(S_t^{opt}, A_t^{opt})$ by recursively discounting future utilities and weighting them by a possible learning rate $\alpha$.
\begin{equation*}
Q\left(S_t^{opt}, A_t^{opt}\right)+\alpha\left[R_t^{LAT}+\gamma \max _{\alpha}\left(Q\left(S^{opt}_{t+1}, \alpha\right)-Q\left(S_t^{opt}, A_t^{opt}\right)\right)\right]
\end{equation*}
\noindent Here $S_t^{opt}$ is the encoded state of the optimizer which is similar to the one of the scheduler:
\begin{equation*}
    S_t^{sche} = S_t^{opt}=\left[ B P_t^{sche},  BR_t^{sche},  QEP_t^{opt},  JD_t^{opt}\right]
\end{equation*}

In the above two equations, the parameter $\gamma$ is the discount factor that weighs the contribution of short-term vs. long-term rewards. Adjusting the value of $\gamma$ will diminish or increase the contribution of future rewards. The parameter $\alpha$ is the learning rate or step size. This simply determines to what extent newly acquired information overrides old information: a low learning rate implies that new information should be treated skeptically and may be appropriate when a workload is mostly stable but contains some outliers. A high learning rate implies that new information is more fully trusted and may be appropriate when query workloads smoothly change over time. Since the above two are recursive equations, they start with making arbitrary assumptions for all $Q$-values. However, as more experience is collected through the execution of incoming queries, the network likely converges to the optimal policy~\cite{dqn}.

%% file: experiments.tex
\begin{figure*}[t]
  \centering
  \begin{subfigure}{0.32\textwidth}
    \includegraphics[width=\textwidth]{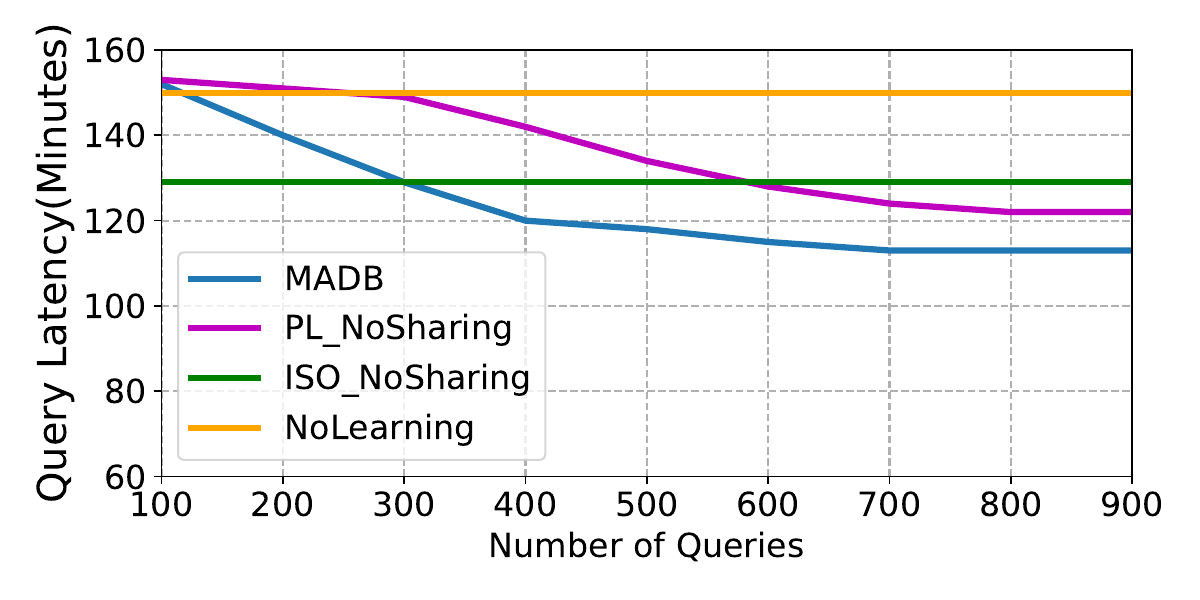}
    \caption{Overall performance improvemet}
    \label{fig:qet_pl}
  \end{subfigure}  
      \begin{subfigure}{0.32\textwidth}
        \includegraphics[width=\textwidth]{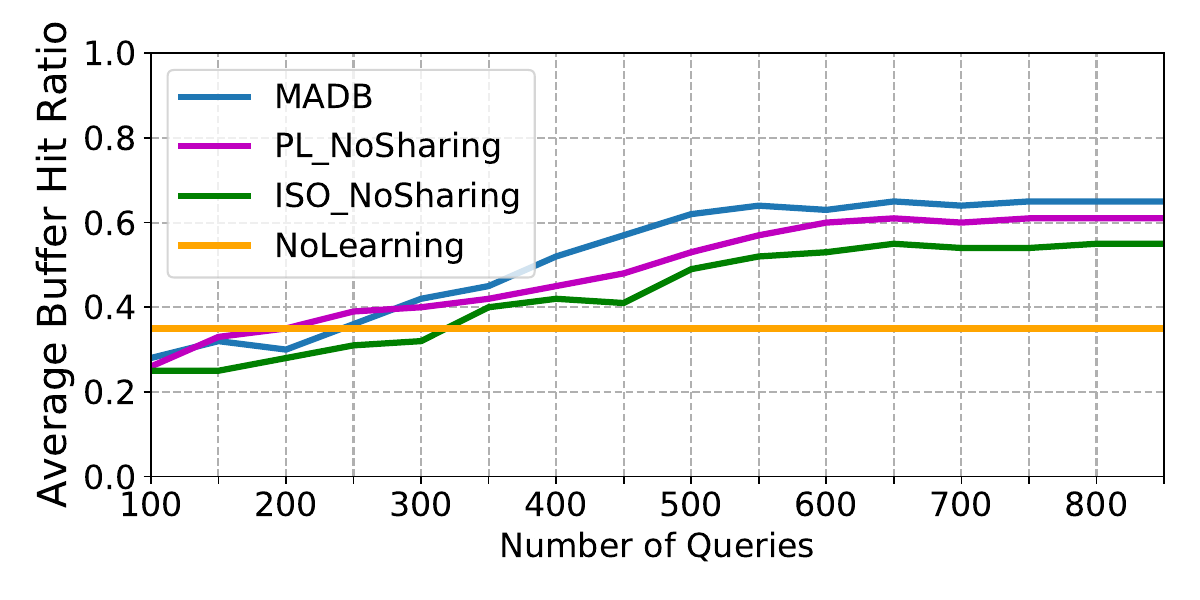}
        \caption{Cache utilization improvement }
        \label{fig:scheduler}
      \end{subfigure}
          \begin{subfigure}{0.32\textwidth}
        \includegraphics[width=\textwidth]{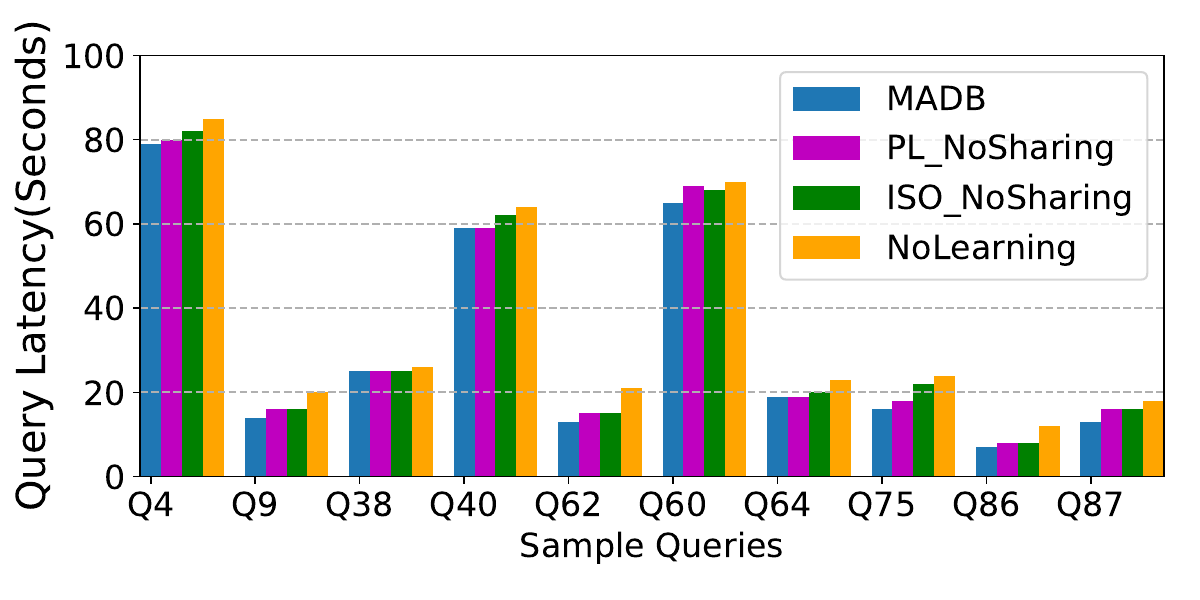}
        \caption { Query plan improvement }
        \label{fig:optimizer}
      \end{subfigure}
  \caption{MADB performance on unseen query templates}
  \label{fig:whole}
\end{figure*}

\section{Preliminary Experiments}\label{t:preliminary}
We present preliminary results demonstrating that MADB with cooperative learned components can improve the overall query performance compared with alternative techniques. 

\noindent{\bf Experimental Setup} Our experimental analysis uses the TPC-DS benchmark~\cite{nambiar2006making}. We deployed a 49GB database on a single node server with four cores and 32GB of RAM. Our database engine was  PostgreSQL~\cite{url-postgres} with a 4GB shared buffer pool\footnote{PostgreSQL was configured to bypass the OS filesystem cache.}. 
 Our learned agents (scheduler and optimizer) are implemented in Keras\cite{keras} using a fully-connected neural network with two hidden layers, each with 128 neurons. We use an adaptive learning rate optimization approach~\cite{slos} and the mean squared error as our loss function. We generate random query instances from the 99 TPC-DS templates. Our agents are trained using $900$  query instances generated from 79 training TPC-DS templates. The testing set includes $100$ instances generated from the remaining 20 TPC-DS templates.

\noindent{\bf MADB Performance} We implemented a proof-of-concept prototype of \emph{MADB} and studied its effectiveness. It uses an  Independent Learning (IL) approach where the scheduler and optimizer are trained in parallel as queries are scheduled and executed and share information about their environments with each other as described in Section~\ref{s:coop}. 

We compare {MADB} to three other approaches. {\tt PL\_NoSharing} includes two agents that are trained in parallel but do not share information. In {\tt ISO\_NoSharing} the agents are trained offline, in isolation from each other (but on the same training workload), and the final trained models are used to schedule and execute the testing queries. {\tt NoLearning} uses no learned components: it combines a first-come-first-served scheduler and PostgresSQL's optimizer.

 Figure \ref{fig:qet_pl} demonstrates the impact of parallel training and information sharing. Here, we increase the size of our training queries from 100 to 900, and each point shows the total time to schedule and execute our 100 testing queries. As MADB is trained on more queries, the execution time of the unseen testing queries improves. MADB also outperforms all other approaches. {\tt PL\_NoSharing} is the second most performing approach but still follows behind MADB by 7\%, demonstrating that state sharing between agents is as important as training concurrently in the same environment. {\tt ISO\_NoSharing} offers a slower execution time (12\% lower than MADB) since the agents are trained in isolation (and do not share any information), and {\tt NoLearning} has the worse performance (24\% slower than MADB). 

\noindent{\bf Per Component Performance}
Next, we studied each learned agent's performance (as defined by its own reward function) and how they perform when trained in a multi-agent environment. Figure \ref{fig:scheduler} shows how the learned scheduler performs as we increase the number of training queries. We measure the average buffer hit ratio of $100$ test queries in unseen templates while increasing our training set. {We observe that the buffer hit ratio in MADB is higher than the alternative approaches.  On average, in MADB, 65\% of the data may be retrieved from the cache rather than the disk, which is 6\% higher than the cache utilization of {\tt PL\_NoSharing},   15\% higher than by {\tt ISO\_NoSharing} and 46\% higher than  {\tt NoLearning}.}

Figure \ref{fig:optimizer} illustrates the latency of 10 randomly selected test queries. The execution plans generated by MADB consistently offer lower latency. Specifically, MADB plans require (on average)  9\% less time than the plans of the PL\_NoSharing approach, 11\% less time than the optimizer's query plan under the ISO\_NoSharing framework, and over 32\% on average less time than the non-intelligent framework NoLearning.

Our results, although preliminary, demonstrate that learned database components trained and executed in a cooperative setting as independent learners can lead to significant performance improvement.

\begin{figure}
  \centering
  \includegraphics[width=0.45\textwidth]{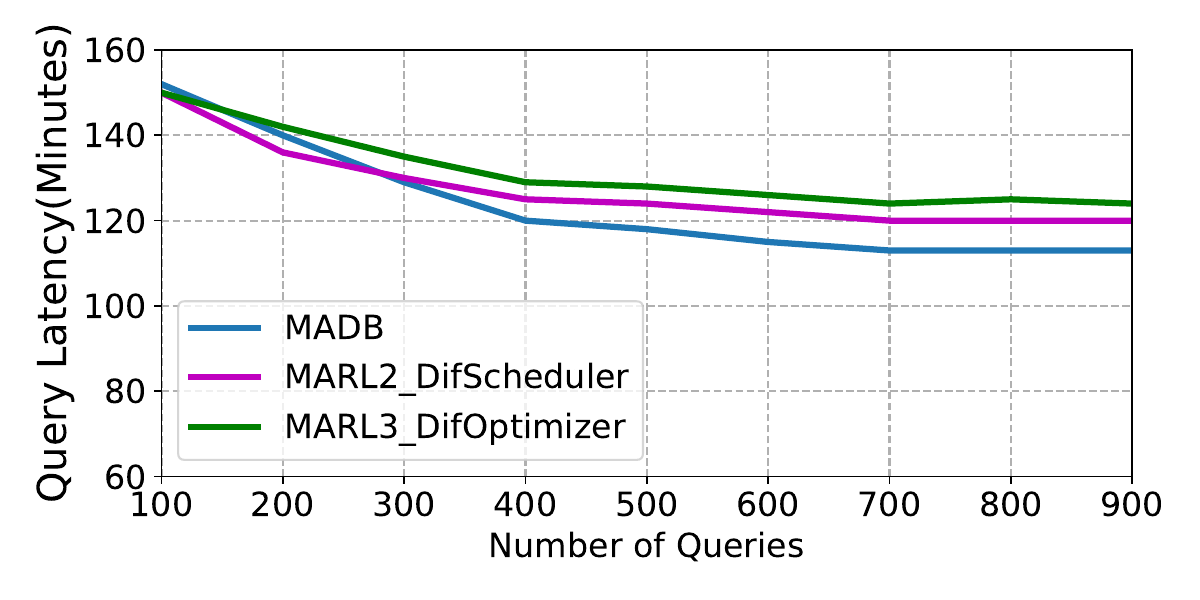}
  \caption{Performance of different sharing information}   \label{fig:qet_share}
\end{figure}

{\bf Impact of Information Sharing}
Lastly, we experimented with our agents sharing different types of information. \\
{\tt MARL2\_DifScheduler} does not receive data block information (the probability of each data block being accessed by a candidate query), but its input includes which tables participate in the query operation.  {\tt MARL3\_DifOptimizer} no longer observes the most recent query execution plan and instead simply records this execution plan's cost. Figure \ref{fig:qet_share} shows that the information shared and observed by the agents in  MADB can maximize the system's performance as more queries are executed since it outperforms by $6\%$ {on average} the different scheduler and $12\%$ {on average} the different optimizer, respectively. 

%% file: conclusion.tex
\section{Conclusion}
\label{sec:conclusion}
We have presented MADB, a multi-agent database system that incorporates a learned query scheduler and a learned query optimizer. Both learned agents in MADB are trained using a cooperative multi-agent reinforcement learning approach, where the agents exchanging information  with the common goal of reducing the  query execution time.  MADB offers substantial improvements over non-cooperative multi-agent learning systems as well as non-learned approaches. This demonstrates that cooperatively integrating multiple learned database components has the potential to identify optimization opportunities that would have been missed by the naive integration of learned components.